\documentclass[aps,prl,twocolumn,superscriptaddress]{revtex4}
\usepackage{epsfig,amssymb,amsmath,amsthm,amsfonts,amsbsy,mathrsfs}
\usepackage{graphicx}
\usepackage{color}
\usepackage{tikz}
\usetikzlibrary{arrows,shapes,chains}
\newcommand*\circled[1]{\tikz[baseline=(char.base)]{
\node[shape=circle,draw,inner sep=0.5pt] (char) {#1};}}

\definecolor{cream}{RGB}{222,217,201}

\begin{document}
\title{Attraction tames two-dimensional melting: from continuous to discontinuous transitions}
\author{Yan-Wei Li}
\affiliation{Division of Physics and Applied Physics, School of Physical and
Mathematical Sciences, Nanyang Technological University, Singapore 637371, Singapore}
\author{Massimo Pica Ciamarra}
\email{massimo@ntu.edu.sg}
\affiliation{Division of Physics and Applied Physics, School of Physical and
Mathematical Sciences, Nanyang Technological University, Singapore 637371, Singapore}
\affiliation{
CNR--SPIN, Dipartimento di Scienze Fisiche,
Universit\`a di Napoli Federico II, I-80126, Napoli, Italy
}
\date{\today}

\begin{abstract}
Two-dimensional systems may admit a hexatic phase and hexatic-liquid transitions of different natures. The determination of their phase diagrams proved challenging,  and indeed those of hard-disks, hard regular polygons, and inverse power-law potentials, have been only recently clarified.
In this context, the role of attractive forces is currently speculative, despite their prevalence at both the molecular and colloidal scale.
Here we demonstrate, via numerical simulations, that attraction promotes a discontinuous melting scenario with no hexatic phase.
At high-temperature, Lennard-Jones particles and attractive polygons follow the shape-dominated melting scenario observed in hard-disks and hard polygons, respectively. Conversely, all systems melt via a first-order transition with no hexatic phase at low temperature, where attractive forces dominate. The intermediate temperature melting scenario is shape-dependent.
Our results suggest that, in colloidal experiments, the tunability of the strength of the attractive forces allows for the observation of different melting scenario in the same system.
\end{abstract}
\maketitle

Two-dimensional (2D) systems with short-range interactions melt either via a first-order solid/liquid transformation or via a two-step process with subsequent solid/hexatic and hexatic/liquid transitions.
The two-step scenario may further follow the Kosterlitz-Thouless-Halperin-Nelson-Young (KTNHY) paradigm~\cite{KT, HN, Y}, with continuous solid-hexatic and hexatic-liquid transitions, or the mixed one~\cite{Krauth2011}, where a discontinuous hexatic-liquid transition follows a continuous solid-hexatic one.
The possibly enormous value of the hexatic correlation length makes it difficult to ascertain which of the above melting scenarios a system follows. However, the increase in computational power and the development of novel algorithms, and careful experiments, allowed to make progresses in recent years. 
For instance, it is now ascertained~\cite{Krauth2011, Experiment_harddisc} that, in hard disks, a discontinuous hexatic-liquid transition follows a continuous solid-hexatic one.
In hard regular polygons, the melting transition depends on the number of edges, e.g., hexagons and squares following the KTHNY melting scenario and pentagons the first-order one~\cite{Glotzer}.
The melting scenario of 2D systems interacting via power-law potentials~\cite{Grimes1979, Keim2007} has been demonstrated to depend on the stiffness of the interaction~\cite{Krauth2015}.
In these recently settled cases, density drives the melting transition, and temperature plays no role as the interaction potentials lack an energy scale.

At both the molecular and colloidal scale, attractive forces are prevalent, and the phase behavior is both temperature and density-dependent. 
The effect of attractive forces on 2D melting remains, however, controversial. 
Indeed, Nelson noticed that attraction may lead to a variety of phase diagrams, illustrating possible scenarios with the hexatic phase occurring in an intermediate temperature range~\cite{Nelson1979,Nelson_book}, for Lennard-Jones (LJ) particles.
This would imply that a weak attraction promotes the hexatic phase, while a strong one suppresses it. 
In attractive systems the existence of the hexatic phase is controversial, as this phase has been observed in some studies~\cite{Wierschem2011, Hayato_EPL}, but not in others~\cite{Di_SoftMatter,Bo_Prx}.
The complete mapping of the phase diagram of LJ particles is a recent, but still debated, achievement~\cite{Hajibabaei2019}; indeed, at high-temperature, where attractive forces are negligible, LJ particles have been found not to follow the melting scenario of $1/r^{12}$~\cite{Krauth2015,Hajibabaei2019} ones.

Here we demonstrate, via the numerical determination of the temperature-density phase diagram of attractive hexagons, pentagons, squares, and LJ point particles, that attraction universally influences the melting scenario by suppressing the hexatic phase and promoting discontinuous transitions.

We simulate attractive hexagons ($N=48071$), pentagons ($N=20449$) and squares ($N=20521$), as well as Lennard-Jones point particles ($N=318^2$), under periodic boundary conditions, in the canonical ensemble using the GPU-accelerated GALAMOST package~\cite{Galamost}.
We construct the extended polygons by lumping together LJ point particles equally spaced along the perimeter, as shown in Fig.~S1~\cite{SM}.
The resulting short-ranged attractive interaction, detailed in the Supplemental Material~\cite{SM}, allows estimating the size of the polygonal particles and their interaction energy scale, we adopt as our units of length and energy, respectively.
For the considered state points and interaction, the values of $N$ we consider are large enough for finite-size effects to be negligible, as we prove in Fig.~S3 in Supplemental Material~\cite{SM}.
We verify thermal equilibration by ascertaining that the same final state is reached in simulations starting from a liquid-like configuration and an ordered one, as illustrated in Fig. S2~\cite{SM}.

\begin{figure}[tb]
 \centering
 \includegraphics[width=0.48\textwidth]{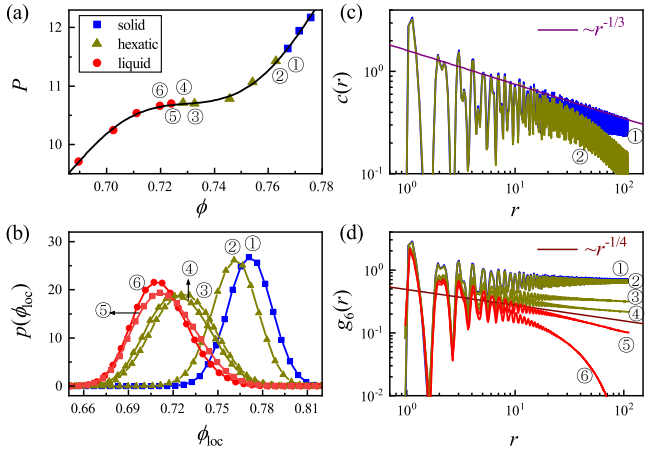}
  \caption{
High-temperature melting of attractive hexagons
(a) Equation of state at $T$=1.40. Different symbols correspond to different phases, as illustrated in the legend. The black line is a fifth-order polynomial fit. (b) The local density histograms, (c) the translational correlation function $c(r)$ and (d) the bond-orientational correlation function $g_{6}(r)$ for different densities, as indicated in (a).
\label{fig:LJKTHNY}
}
\end{figure}

\emph{Attractive hexagons at high temperatures. -- } We begin by reporting results for the determination of the phase diagram of attractive hexagons. 
At high temperature, the pressure of attractive hexagons increases monotonically with the density, as illustrated in Fig.~\ref{fig:LJKTHNY}(a).
We associate to each particle a local density defined as $\rho(\vec{r}_i)=\frac{\sum_{j=1}^{N}H(r_c-|\vec{r}_{i}-\vec{r}_{j}|)}{\pi r_c^2}$, where $H$ is the Heaviside step function, $r_c=50$. Results are robust with respect to choice of $r_c$, unless it becomes very small, or of the order of the system size. 
Figure~\ref{fig:LJKTHNY}(b) shows the distribution of the local density, which is always unimodal. 
The density dependence of the pressure and of the local density distribution exclude the presence of a discontinuous transition with a coexistence phase.
We identify the different pure phases investigating the spatial decay of the correlation function of the translational, $c(r)$, and of the bond-orientational order, $g_6(r)$. The translational correlation function is $c(r=|\vec{r}_{i}-\vec{r}_{j}|)=e^{i {\vec G} \cdot (\vec{r}_{i}-\vec{r}_{j})}$, where $\vec{G}$ is one of the first Bragg peaks, identified by the static structure factor~\cite{Krauth2011, Massimo_PRE2019}.
The  bond-orientational correlation function is $g_{k}(r=|\vec{r}_{i}-\vec{r}_{j}|)=\langle\psi_{k}(\vec{r}_{i})\psi_{k}^{*}(\vec{r}_{j})\rangle$, where $\psi_{k}(\vec{r}_i)$ is the bond-orientational order parameter of particle $i$, defined as $\psi_{k}(\vec{r}_i)=\frac{1}{n}\sum_{m=1}^{n}\exp(ik\theta_{m}^{i})$. Here, $n$ is the number of nearest neighbors of the particle and $\theta_{m}^{i}$ is the angle between $(\vec{r}_{m}-\vec{r}_{i})$ and a fixed arbitrary axis. The value of $k$ reflects the rotational symmetry of the crystal structure: $k = 4$ for squares, $k = 6$ for the other particles.

At high density, the system is in the solid phase.
Consistently, we observe the translational correlation function to decay as $c(r)\propto r^{-\eta}$ with $\eta \le 1/3$, a consequence of the Mermin-Wagner theorem~\cite{Mermin}, and the bond-orientational correlation function to reach a constant, as illustrated in Figs.~\ref{fig:LJKTHNY}(c) and \ref{fig:LJKTHNY}(d) for state point \circled{1}. At lower density, $c(r)$ decays exponentially, while $g_{6}(r)$ has a power-law decay, $g_{6}(r)\propto r^{-\eta_{6}}$ with $\eta_{6}<1/4$. This occurs, for instance, at state points \circled{2}-\circled{4}, and indicates that the system is in the hexatic phase.
Further lowering the density, the system enters the liquid phase, where both correlation functions decay exponentially.

These findings demonstrate that, at high temperature, LJ hexagons follow the KTHNY scenario~\cite{KT, HN, Y}. This result agrees with a previous investigation of the melting transition of hard hexagons~\cite{Glotzer}, the role of attractive forces being negligible at high temperatures.

\begin{figure}[!t]
 \centering
 \includegraphics[angle=0,width=0.5\textwidth]{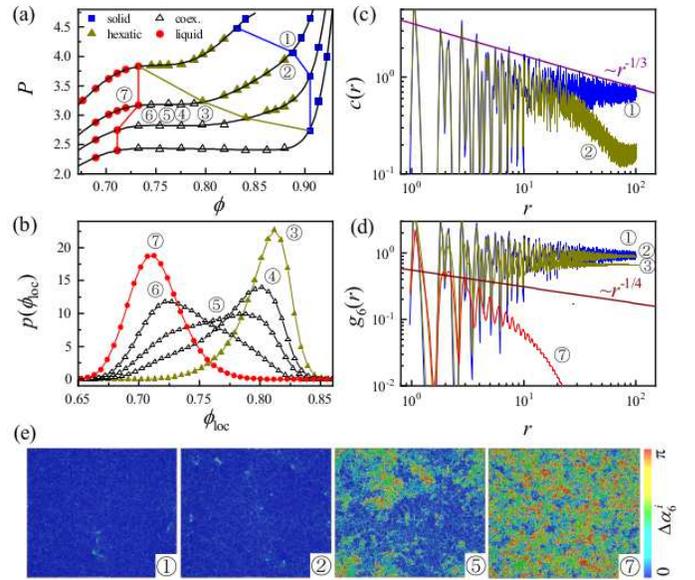}
 \caption{
Intermediate-temperature melting of attractive hexagons.
(a) Equation of state at $T$=0.60, 0.53, 0.49 and 0.46, from top to bottom. Different symbols correspond to different phases, as illustrated in the legend. The black lines are from polynomial fits. Phase boundaries are marked by the red, dark yellow and blue lines. (b) Local density distributions, (c) translational and (d) bond-orientational correlation functions, and (e) snapshots of the system at $T$=0.53, for different densities. 
In (e) each hexagon is colour coded according to the angle between its local bond-orientational parameter, and the global one.
\label{fig:ljHT}
}
\end{figure}
\emph{Attractive hexagons at intermediate  temperature. --} As the temperature decreases the equation of state of attractive hexagons flattens, in a range of densities, and develops a Mayer-Wood~\cite{Mayer_wood} loop for $T \lesssim 0.53$, as illustrated in Fig.~\ref{fig:ljHT}(a). 
Since pressure loops are induced by the interfacial free energy of coexisting phases~\cite{Binder1982, Krauth2011}, this indicates the presence of a first-order transition.
Within the coexisting region, the distribution of the local density becomes extremely broad and well described by the superposition of two Gaussian functions, as shown in Fig.~\ref{fig:ljHT}(b). 
Besides, the distribution becomes system-size dependent, with a bimodal character more apparent in larger systems, as we show in Fig. S3~\cite{SM}.
These findings further support the presence of coexisting phases. 

We determine the coexistence boundaries via the Maxwell construction with the pressure curve fitted by either a fifth- or a tenth-order polynomial.
Outside of the coexistence region, we identify the pure phases investigating the translational and the bond-orientational correlation functions, as summarized in Figs.~\ref{fig:ljHT}(c) and \ref{fig:ljHT}(d).
We observe the solid phase (e.g., \circled{1}), where $c(r)$ decays algebraically and $g_{6}(r)$ is extended, the hexatic phase (e.g., \circled{2}), where $c(r)$ decays exponentially and $g_{6}(r)$ is extended, and the liquid phase (e.g., \circled{7}) where both correlation functions decay exponentially. 
These results indicate that, at intermediate temperatures, attractive hexagons follows the mixed melting scenario with a continuous solid-hexatic transition anticipating a discontinuous hexatic-liquid transition.

We visualize the different phases by colour-coding each particle according to the angle $\Delta \alpha_k^i$ between the global $\vec{\Psi}_{k} = \frac{1}{N} \sum_i \psi_{k}(\vec{r}_i)$ and the local $\psi_{k}(\vec{r}_i)$ bond-orientational parameters, $\psi_{k}(\vec{r}_i)\cdot \vec{\Psi}_{k}^* = |\psi_{k}(\vec{r}_i)||\vec{\Psi}_{k}^*|\cos(\Delta \alpha_k^i)$.
In the solid and hexatic phase, the long-range or quasi-long-range nature of the bond-orientational order leads to snapshots with a uniform colour,  as in Fig.~\ref{fig:ljHT}(e) \circled{1} and \circled{2}.
In the liquid phase, Fig.~\ref{fig:ljHT}(e) \circled{7}, the snapshot appears almost randomly coloured, due to the short-range of the bond-orientational order.
In the coexistence phase, Fig.~\ref{fig:ljHT}(e) \circled{5}, the coexistence of hexatic and liquid phases lead to that of regions of uniform colour and regions randomly coloured.

\begin{figure}[!t]
 \centering
 \includegraphics[angle=0,width=0.48\textwidth]{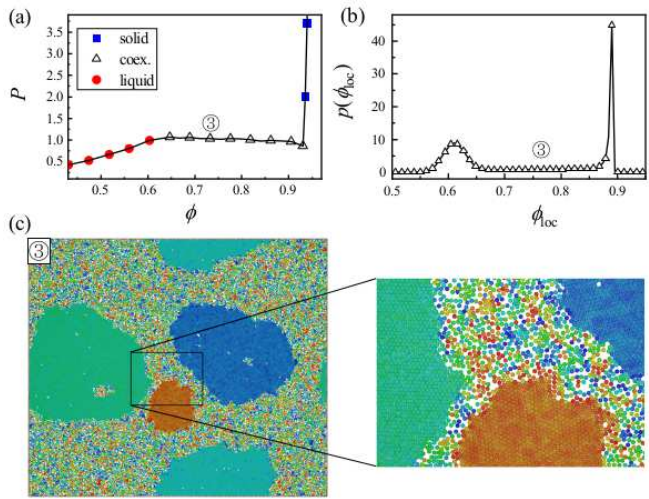}
 \caption{Low-temperature melting of attractive hexagons.
(a) Equation of state at $T=0.35$. The black line is a guide to the eyes. (b) Local density distributions at a density value withing the coexistence region. (c) snapshot of the whole system (left), and enlargement of part of it (right), at the corresponding value of the densiy. The colour code is as in Fig.~\ref{fig:ljHT}(e).
\label{fig:ljLT}
}
\end{figure}

\emph{Attractive hexagons at low  temperature. --} As the temperature decreases, the coexistence region widens, and the hexatic phase shrinks, as apparent in Fig.~\ref{fig:ljHT}(a). At low enough temperatures, therefore, melting occurs via a first-order liquid-solid transition with no hexatic. The pressure loop and the bimodal character of the local density distribution within the coexistence region, which we illustrate in Fig.~\ref{fig:ljLT}, confirms that the system undergoes a discontinuous transition at low temperature. 

Furthermore, snapshots of the system indicate that the coexisting phases are of solid and liquid type, as in Fig.~\ref{fig:ljLT}(c). 
We can, therefore, exclude an intermediate hexatic phase, further supporting a first-order melting scenario at low temperature.

\begin{figure}[!t]
 \centering
 \includegraphics[angle=0,width=0.5\textwidth]{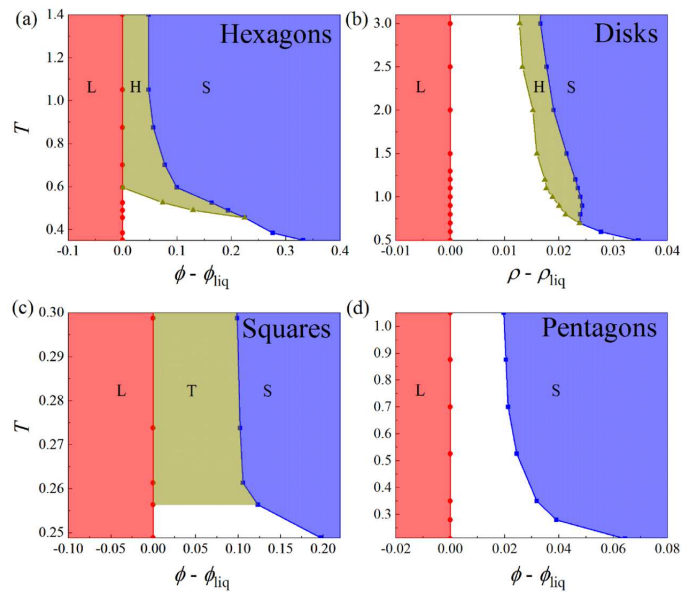}
 \caption{
Phase diagram of attractive polygons and LJ disks. 
The phase diagrams for (a) hexagons, (c) squares and (d) pentagons are plotted in the $T$-$\phi-\phi_{\rm liq}$ plane and the one for (b) disks is in $T$-$\rho-\rho_{\rm liq}$ plane, where $\phi_{\rm liq}$ ($\rho_{\rm liq}$) is the highest area fraction (number density) of the liquid phase.
At each investigated temperature, we mark with symbols the estimated phase boundaries. 
Colours are used to distinguish the pure phases, liquid (L), solid (S), hexatic (H) and tetratic (T). Coexistence regions, including hexatic-liquid and solid-liquid coexistence, are white.
The corresponding phase diagrams in the $T$-$\phi$ (or $T$-$\rho$) plane are in Fig.~S6~\cite{SM}.
\label{fig:pd}
}
\end{figure}

\emph{Shape and temperature dependence of the melting scenario. --} The phase diagram of Fig.~\ref{fig:pd}(a) summarizes the results we have obtained so far: for attractive hexagons, melting is of KTHNY type at high temperature and becomes first of mixed type and then first-order as the temperature decreases. The system has no liquid-gas transition, nor a solid-solid transition, as the attraction range, we determine in Table \uppercase\expandafter{\romannumeral1} in the Supplemental Material~\cite{SM}, is small but not much smaller than the particle size. This suppresses the liquid-gas critical point~\cite{Frenkel_PRL1994} without promoting a solid-solid transition~\cite{Frenkel_PRL1994, Frenkel_1995, Marjolein2002}. In the solid, the body-orientation of the hexagonal particles is long-ranged, thus excluding the presence of a plastic-crystal phase.

We investigate the universality of the role of attraction in the 2D melting by determining the phase diagram for different particle shapes: squares, pentagons and LJ point particles (disks). 
Squares crystallize in the square lattice, all other shapes in the hexagonal one. 
For pentagons, shape-frustration is not able to inhibit crystallization at the lowest temperature ($T=0.19$) and the highest density ($\phi=0.854$) we studied.
Details on the phase determination are in Figs.~S4 and S5~\cite{SM}, for squares and pentagons, and elsewhere for disks~\cite{Future}.
The resulting phase diagrams are in Figs.~\ref{fig:pd}(b)-(d).

At high temperature, the polygonal particles follow the melting scenario previously reported for hard particles~\cite{Glotzer}, KTHNY for squares and first-order for pentagons. LJ disks follow the mixed scenario as $r^{-12}$ particles~\cite{Krauth2015}.
As in hexagons, in both squares and pentagons, the liquid-gas transition is suppressed, and no plastic-crystal phase occurs. In LJ disks, the liquid-gas critical point occurs at low-temperature and low-density, well outside the parameter space we have investigated.

Regardless of the high-temperature behaviour, melting always occurs via a first-order solid-liquid transition at low temperature, and the coexistence region broadens as the temperature decreases.
These findings imply that particles' shapes fix the melting scenario at high temperature, attractive forces at low temperatures. 
At intermediate temperatures, conversely, both shape and attraction may be relevant.
In disks, the hexatic phase disappears as the temperature decreases, making the melting transition first-order. 
In squares, the equation of state within the tetratic region becomes flat as the temperature decreases without the coexistence region shrinking, as we illustrate in Fig. S4~\cite{SM}. Hence, no intermediate mixed scenario separates the high-temperature continuous melting and the low-temperature discontinuous one. In this respect, squares differ from hexagons and disks.
In pentagons, the transition is always first-order. 

These results consistently demonstrate that attraction influences the melting scenario of 2D systems by promoting the emergence of a coexistence region, if this is not already present in the high-temperature limit, as well as widening it. 
The widening of the coexisting region leads to the disappearance of the hexatic/tetratic phase, and hence to a first-order melting transition.

\emph{Conclusions. --}
Different system properties affect the melting scenario in 2D~\cite{Krauth2015, Glotzer, OurPaper, John_Russo, ningxu}.
In this context, the influence of attractive forces was unclear, despite their prevalence at both the molecular and colloidal scale.
We have found that attractive forces induce a discontinuous transition and widen the coexistence region at the expense of the hexatic phase, making the low-temperature melting transition first-order.
Hence, attractive forces never induces the hexatic phase or widens the hexatic region, at variance with previous speculations~\cite{Nelson1979,Nelson_book}.
We suggests that attractive forces always promote the discontinuous transition, as we demonstrated this to occur in systems which melt according to different scenarios at high temperature.

Theoretically, our results suggest that the dislocation core energy, $E_c$, is suppressed at low temperatures, in the presence of attractive forces. 
Conversely, at high-temperature $E_c/k_bT \gg 1$, and a continuous two-step melting scenario may occur, according to the KNTHY theory. 
We are looking forward to the experimental investigation of our predictions in colloidal systems, where the tuning of the strength of the attractive forces, e.g. via the depletion interaction, should allow for the observation of different melting scenario in the same system, e.g. colloidal hexagonal-shaped particles.

The interparticle interaction of our polygonal particles inherits their discrete rotational symmetry, as the attraction range is small compared to the particle size, as we detail in the Supplemental Material~\cite{SM}. As the attraction range increases, this discrete rotational symmetry vanishes and the interaction becomes more rotationally symmetric. Hence, while we have not explicitly investigated the role of the attraction range on the phase behaviour, we anticipate that on increasing the attraction range the phase diagrams of the polygonal particles evolve towards that of the LJ point particles.

We acknowledge support from the Singapore Ministry of Education through the Academic Research Fund MOE2017-T2-1-066 (S), and are grateful to the National Supercomputing Centre (NSCC) of Singapore for providing computational resources. We thank Joyjit Chattoraj for helpful discussions

\bibliographystyle{apsrev4-1}
%

\pagebreak
~\newpage
\onecolumngrid
\section{Interparticle interaction\label{sec:pot}}
\begin{figure*}[b!]
 \centering
 \includegraphics*[angle=0,width=0.7\textwidth]{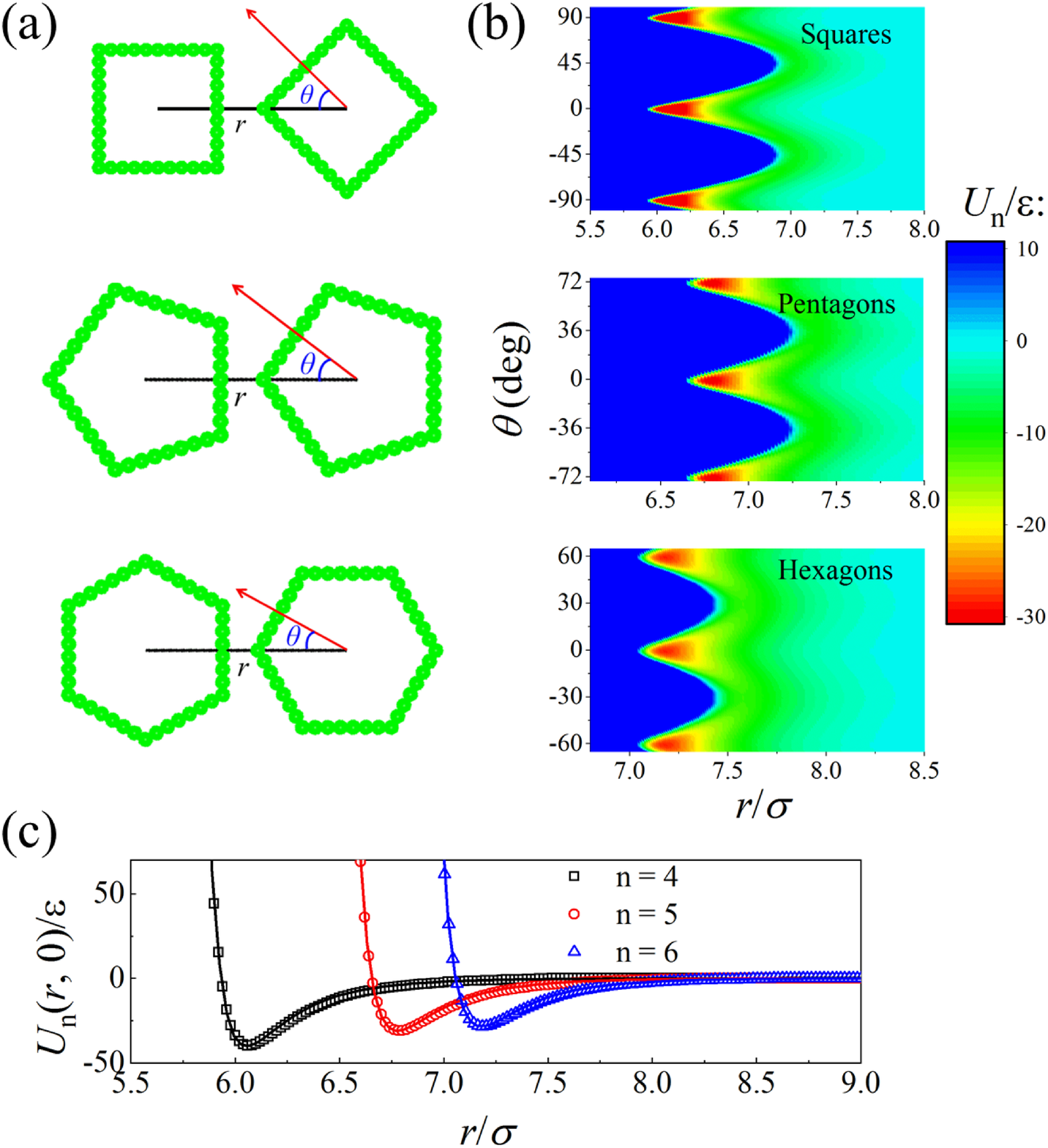}
\caption{Polygonal particles' interaction. (a) We construct the polygonal particles by lumping together point particles along their perimeters. Point particles of different polygons interact via a LJ potential. 
(b) Energy of interaction of polygonal particles, as a function of their separation $r$ and relative orientation, $\theta$. 
(c) Dependence of the interaction energy between polygonal particles on the inter-particle distance, for $\theta = 0$. Lines are fits to Eq.~\ref{eq:lj}.
\label{fig:pot}}
\end{figure*}

We construct the polygons by lumping together $N_d$ point particles equally spaced along the perimeter, as illustrated in Fig.~\ref{fig:pot}(a).
Point particles of different polygons interact via a LJ potential, $V_{\rm LJ}(r) = 4\epsilon \left[ \left(\frac{\sigma}{r}\right)^{12}- \left(\frac{\sigma}{r}\right)^{6} + c\right]$ for $r \leq r_c = 2.5\sigma$, $V_{\rm LJ}(r) = 0$ otherwise, where the constant $c$ enforces $V_{\rm LJ}(r_c) = 0$. The radius of the circumcircle is $5\sqrt{2}/2\sigma$ for squares, pentagons and hexagons.

The interaction energy $U_{\rm n}(r,\theta)$ of two polygonal particles with ${\rm n}$ sides is the sum of the interactions of their constituent point particles. 
This interaction depends on the separation between the polygons, $r$, and their relative orientation, $\theta$. 
The angular dependence inherits the rotationally symmetry of the particles, $U_{\rm n}(r,\theta)=U_{\rm n}(r,\theta+\frac{2\pi}{\rm n})$.
For the different particles, we illustrate $U_{\rm n}(r,\theta)$, with the energy expressed in units of $\epsilon$, and the distances in units of $\sigma$, in Fig.~\ref{fig:pot}(b).
For the number of point particles $N_d$ we use to construct the polygons, the interaction energy is de-facto linear in $N_d$. The discrete representation of the polygonal particles is thus not affecting our results.

For $\theta = 0$, $U_{\rm n}(r,\theta= 0)$ is well described by a LJ potential for extended particles, \begin{equation}
U_{\rm n}(r,\theta= 0) =4\epsilon_{\rm n} \left[
\left(\frac{\sigma}{r-d_{\rm n}}\right)^{12}-\left(\frac{\sigma}{r-d_{\rm n}}\right)^{6}
\right],
\label{eq:lj}
\end{equation}
as we illustrate in Fig.~\ref{fig:pot}(c).
Hence, for $\theta = 0$, the interaction potential is zero at $r_0=d_{\rm n}+\sigma$, and reaches its minimum value $\epsilon_{\rm n}$ at $r_{\rm min}=d_{\rm n}+2^{1/6}\sigma$. 
The parameter $\lambda_{\rm n} = \frac{2(r_{\min}-r_0)}{r_0}$ is a measure of the width of the attractive range with respect to the particle size.
We summarized the values of these parameters in Table~\ref{tb:parameters}. 


\begin{table}[t]
    \centering
\begin{tabular}{|l |c |c | c |}
 \hline &~Squares (n=4)~&~Pentagons (n=5)~&~Hexagons (n=6)~\\ \hline $\epsilon_{\rm n}/\epsilon$ & 40.2 & 31.1 & 28.5 \\ 
 \hline $d_{\rm n}/\sigma$ & 4.9 & 5.6 & 6.0   \\
 \hline $N_d$ & 40 & 40 & 42   \\
 \hline $r_0/\sigma$ & 5.9 & 6.6 & 7.0   \\
 \hline $\lambda_{\rm n}$ & 0.042 & 0.037 & 0.035   \\
 \hline
\end{tabular}
\caption{For the polygonal particles, we provide the values of the parameters $\epsilon_{\rm n}$ and $d_{\rm n}$ describing the interparticle interaction for $\theta = 0$, given the number of beads $N_d$ we use to describe the particles, as well as the particle size $r_0$ and the ratio between the attraction range and the particle size, $\lambda_n$.
\label{tb:parameters}
}    
 \end{table}

\section{Units\label{sec:units}}
For LJ disks, we use standard LJ units: $\epsilon$, $\sigma$ and $m$ are our units of energy, distances and masses, where $m$ is the mass of each particle.

For polygonal particles, we use units facilitating comparisons with colloidal-scale experiments, where the depletion interaction controls the strength and the range of the interparticle interaction, and the area fraction measures the density.
The minimum of the interaction energy, $\epsilon_n$, is the unit of energy, and the distance at which the interaction potential between aligned particles is zero, $r_0$, is the unit of length. The unit of mass is the mass of the particle, $mN_d$, with $m$ mass of the point particles used to construct the extended polygons, and $N_d$ their number. The values of these quantities in standard LJ units are in Table~\ref{tb:parameters}. 
Interpreting our unit of length as the diameter of the circle inscribed in the polygonal particles, we evaluate their area $A_n$. We measure the degree of crowding via the area fraction $\phi = \rho A_n$, with $\rho$ number density.

\newpage
\section{Thermal equilibration\label{sec:Equ}}
\begin{figure*}[h!]
 \centering
 \includegraphics*[angle=0,width=0.75\textwidth]{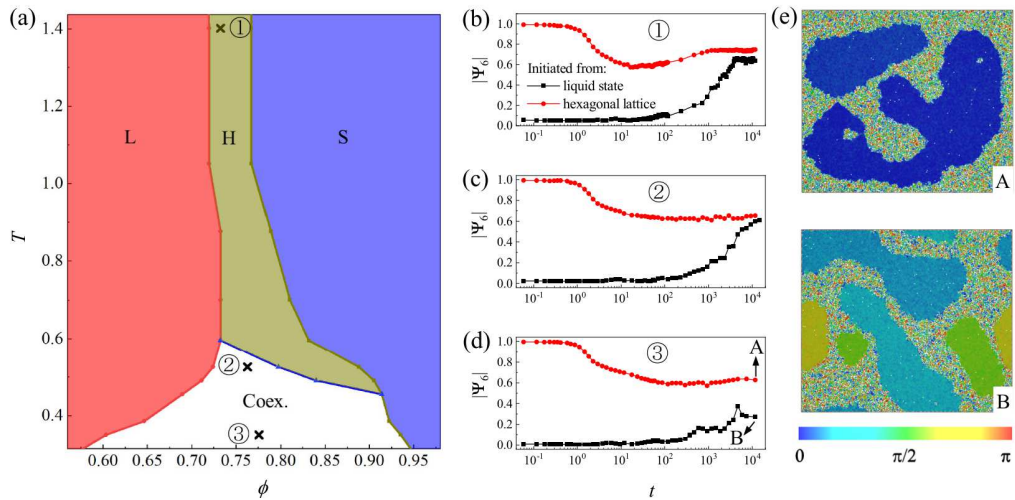}
\caption{
Equilibration of attractive hexagons. (a) phase diagram of hexagons in $T-\rho$ plane. (b), (c), and (d) illustrate the time evolution of the global bond-orientational order parameter for the three state points marked in (a). The initial configuration is either in the liquid phase, or in the solid one.
(e) snapshots of the final configurations reached in (a). 
\label{fig:Equili}}
\end{figure*}

We ensure that our simulations reach thermal equilibrium by checking for the convergence of simulations starting from a liquid-like and a crystalline configuration, at long times. The liquid configuration is prepared by quickly compressing a dilute configuration to the target density. We illustrate this analysis for the hexagonal particles, focusing on three values of the control parameters in the hexatic phase (\circled{1}), the hexatic-liquid (\circled{2}) and the solid-liquid (\circled{3}) coexistence regions (see Fig.~\ref{fig:Equili}(a)).

Regardless of the initial configuration, the global bond-orientational parameter reaches the same asymptotic value at long-times, for state points in the hexatic phase and the hexatic-liquid coexistence regions, as we illustrate in Figs.~\ref{fig:Equili}(b) and \ref{fig:Equili}(c). We do not always find this convergence for state points within the solid-liquid coexistence region, as exemplified in Fig.~\ref{fig:Equili}(d); nevertheless, the system always reaches the solid-liquid coexistence phase, as demonstrated by the snapshots of the final configurations we provide in Fig.~\ref{fig:Equili}(e).
Hence, the different long-time values of the global bond-orientational parameters reflect the different geometrical shapes of the coexisting phases. As the system coarsen, on a time-scale not accessible in our simulations, the observed difference should disappear.

Equilibration runs of at least $t=10^{5}$, for point particles, and $t=10^{4}$, for polygonal systems, have been carried out before collecting data. We have verified that this equilibration time ensured that the system reaches thermal equilibrium at the considered state points.
\newpage

\section{Finite-size effects \label{sec:size}}
We investigate the size dependence of the melting scenario of hexagonal particles at three different temperatures, $T=1.40$, $0.53$, and $0.35$.
At these temperatures, melting follows the KTHNY, the mixed and the first-order scenario, respectively. 
We vary the number of polygons from 
$N=10661$ to $48071$.

\begin{figure}[h!]
 \centering
 \includegraphics[angle=0,width=0.7\textwidth]{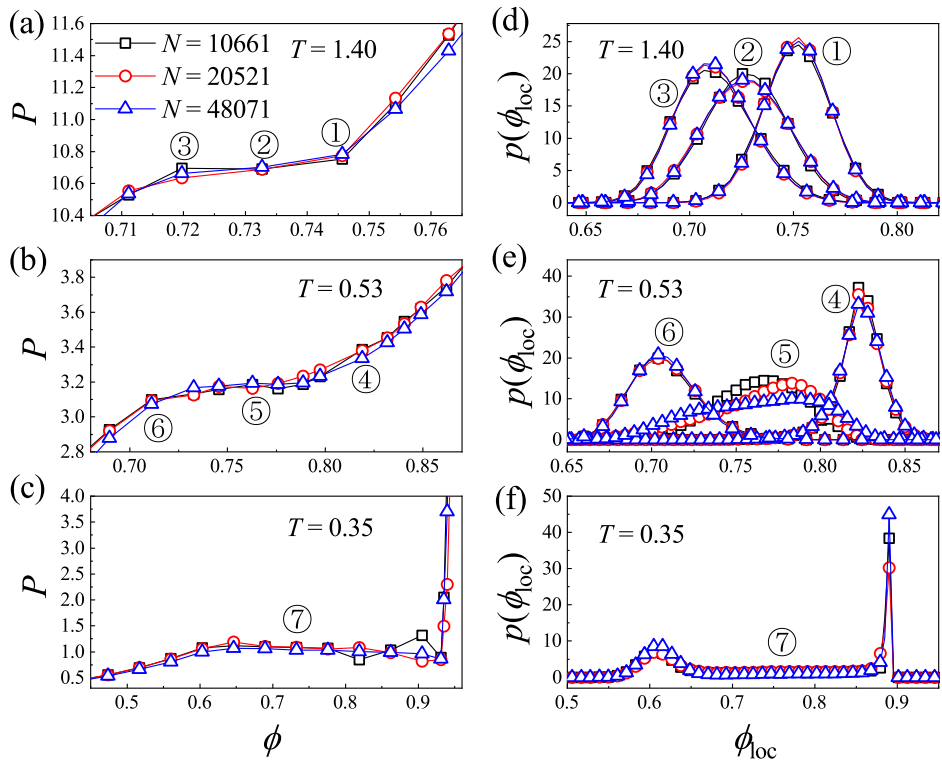}
\caption{Size effects in attractive hexagons.
Equation of state and distribution of the local density at $T=1.40$ ((a) and (d)),  $0.53$ ((b) and (e)), and $0.35$ ((c) and (f)). The different symbols correspond to different system sizes, as indicated in (a). The lines are guides for the eye.
\label{fig:eos}}
\end{figure}

The equation of state may exhibit, in principle,  Mayer-Wood loops across both continuous and discontinuous transition.
Across continuous transition, the amplitude of the loop scales as $N^{-1}$ and the size dependence is lost even in moderate system sizes. 
Indeed, we do not observe any size dependence in the equation of state at high temperature, as we illustrate in Fig.~\ref{fig:eos}(a), the system melting via two continuous transitions.

Across discontinuous transitions, a size dependence commonly occurs as the amplitude of the Mayer-Wood loops scales as $N^{-1/2}$.
We do not observe any size dependence in hexagonal systems even in the presence of discontinuous transition, as illustrated in Figs.~\ref{fig:eos}(b) and \ref{fig:eos}(c), the equation of state being flat in the coexistence region, even for the smallest systems we have investigated. 
This peculiarity makes easy the identification of the coexisting densities.

Consistently with these results, the local density distribution results size-independent at high temperature.
At intermediate temperatures, the system melts via the mixed scenario, and the local density distribution is size-dependent in the hexatic-liquid coexistence region. In particular, the distribution broadens with the system size increases,  suggesting a bimodal behaviour in larger systems.
The size dependence is also weak at $T = 0.35$, where the system melts via a first-order transition, and within the coexistence region, the local density distribution is bimodal.

Squares and pentagons behave similarly to the hexagons, as concern the size dependence. Conversely, we observe the $N^{-1/2}$ size dependence across the liquid-hexatic transition of LJ point particles.

\newpage

\section{Squares: phase determination \label{sec:squ}}
\begin{figure*}[h!]
 \centering
 \includegraphics[angle=0,width=0.85\textwidth]{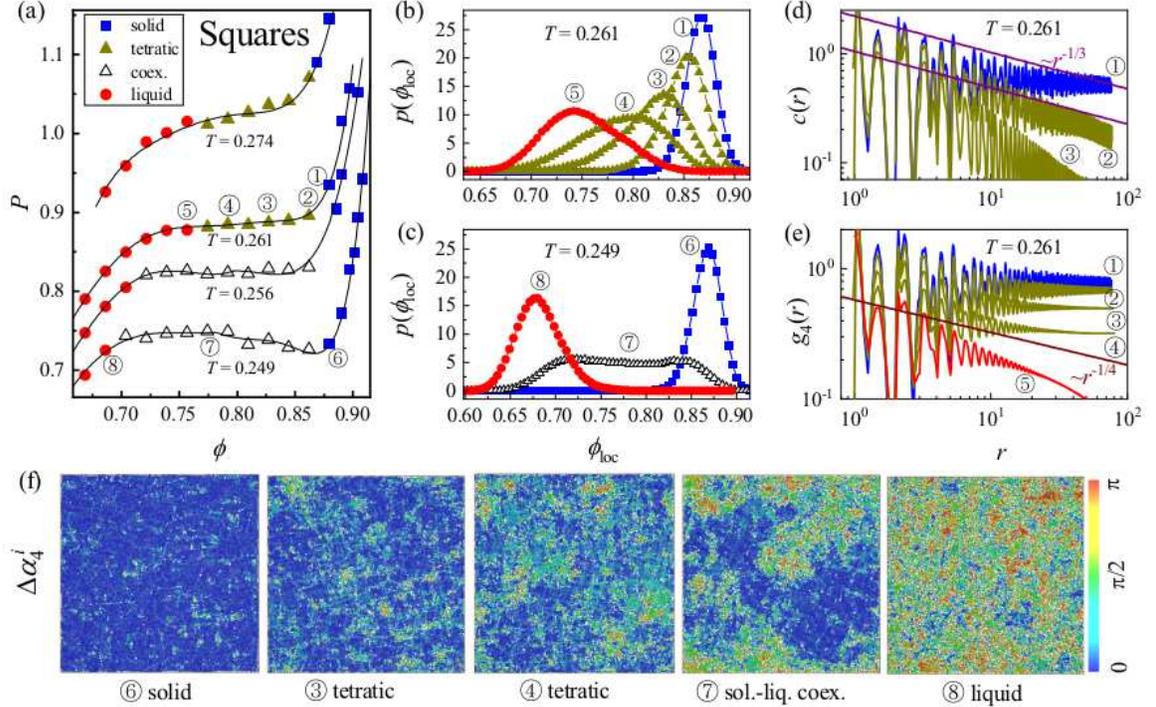}
\caption{
 Melting of attractive squares.
(a) Equation of state of squares, at selected values of the temperature. Different symbols correspond to different phases, as in the legend. Black lines are polynomial fits. 
(b) and (c) illustrate the distribution of the local density at selected state points. (d) and (e) illustrate the decay of the translational and of bond-orientational correlation functions. 
In (f), we illustrate snapshots of the system with particles colour coded according to the angle $\Delta \alpha^i_4$ between their local bond-orientational order parameter, and the global one.
\label{fig:square}}
\end{figure*}

At high-temperature squares melt via continuous transitions as no pressure loop occurs in the equation of state, as in Fig.~\ref{fig:square}(a), $T=0.274$.
As the temperature decreases the pressure flattens, the transition becoming discontinuous below $T \lesssim 0.256$.
Consistently, at high-temperature, the distribution of the local density is unimodal at all densities, while at low temperature, the distribution is bimodal within the coexistence region, as in Figs.~\ref{fig:square}(b) and \ref{fig:square}(c).
We identify the pure phases studying the decay of the correlation functions of the translational and of the bond-orientational order parameters. 
In squares, the bond-orientational correlation function is
$g_{4}(r=|\vec{r}_{i}-\vec{r}_{j}|)=\langle\psi_{4}(\vec{r}_{i})\psi_{4}^{*}(\vec{r}_{j})\rangle$, given the expected square symmetry of the crystal, and the intermediate phase between the crystal and the liquid phase is tetratic, rather than hexatic.

We illustrate the decay of the correlation functions, for the state points \circled{1}--\circled{5} identified in Fig.~\ref{fig:square}(a), in Figs.~\ref{fig:square}(d) and \ref{fig:square}(e). 
We observe the solid phase (e.g., \circled{1}), where $c(r)$ decays algebraically and $g_{4}(r)$ is extended, the tetratic (e.g., \circled{2} and \circled{3}) phase, where $c(r)$ decays exponentially and $g_{4}(r)$ decays algebraically (or extended in the observation window), and the liquid one (e.g., \circled{5}) where both correlation functions decay exponentially.

Snapshots of the system with the colour of the squares reflecting the angle between their local bond-orientational order parameter and the global one, we illustrate in Fig.~\ref{fig:square}(f), support the above identification of the phases. In the solid and tetratic phase, the uniform colour reflects the long-range or quasi-long-range nature of the bond-orientational order. Random colours characterize the liquid phase, as for state point \circled{8}, while in the coexistence phase homogeneously coloured patches coexist with more random ones.

We remark that the temperature dependence of the equation of state in hexagons and square is qualitatively different, although both systems follow the KTNHY scenario at high-temperatures and the first-order one at low temperatures. In hexagons, the hexatic phase gradually shrinks while the coexistence region widens, as the temperature decreases; in squares, conversely, the coexistence phase abruptly replaces the tetratic one. A mixed melting scenario thus occurs in hexagons, but not in squares, at intermediate temperatures.

\newpage
\section{Pentagons: phase determination \label{sec:pent}}
\begin{figure*}[h!]
 \centering
 \includegraphics[angle=0,width=0.85\textwidth]{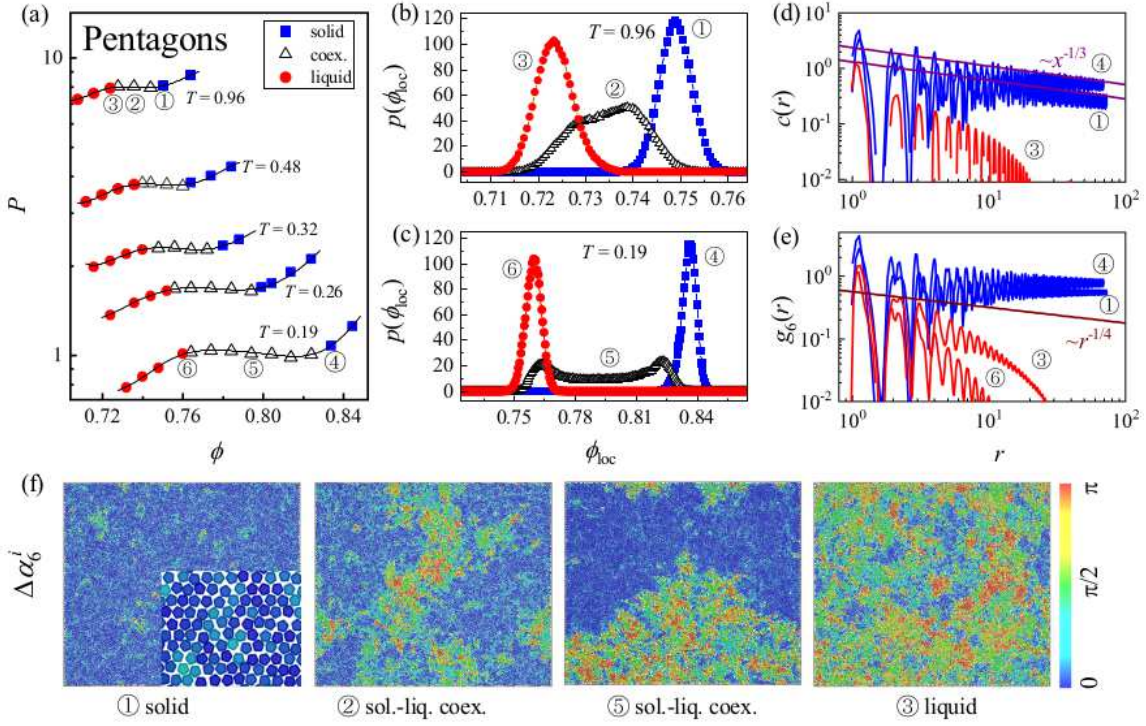}
\caption{
Melting of attractive pentagons.
(a) Equation of state, at selected values of the temperature. Different symbols correspond to different phases, as in the legend. Black lines are from polynomial fits. For selected state points, we illustrate the distribution of the local density in (b) and (c), and the decay of the translational and bond-orientational correlation functions in (d) and (e). In (f), we illustrate snapshots of the system with particles colour coded according to the angle $\Delta \alpha^i_6$ between their local bond-orientational order parameter, and the global one. An enlargement of the solid phase is shown on the bottom right of the left panel in (f).
\label{fig:pentagon}}
\end{figure*}

The equation of state of attractive pentagons reveals a discontinuous melting transition at all temperatures with the coexistence region broadening at low temperature, as in Fig.~\ref{fig:pentagon}(a).
Consistently, at all temperatures, there is a coexistence region where the distribution of the local density is bimodal, as in Figs.~\ref{fig:pentagon}(b) and \ref{fig:pentagon}(c).

We identify the pure phases outside the pressure loop investigating the translational and the bond-orientational correlation functions. Examples are in Figs.~\ref{fig:pentagon}(d) and \ref{fig:pentagon}(e).
We only observe the hexagonal solid phase (e.g., \circled{1} and \circled{4}), where $c(r)$ decays algebraically and $g_{6}(r)$ is extended, and the liquid one (e.g., \circled{3} and \circled{6}) where both correlation functions decay exponentially. Pentagons, therefore, melt via the first-order scenario at all temperatures.

Snapshots of the system, with the particles coloured according to the angle $\Delta \alpha_6^i$ their local bond-orientational parameter form with the global one, confirm this interpretation of the melting scenario, as illustrated in Fig.~\ref{fig:pentagon}(f).

\newpage

\section{Phase diagrams in $T-\phi$ plane\label{sec:pd}}
In the main text, we have presented the phase diagram in the $T$ vs $\phi-\phi_l$ plane for polygonal systems and in the $T$ vs $\rho-\rho_l$ plane for disks, where $\phi_l$ ($\rho_l$) is the highest area fraction (number density) of the pure liquid phase, at each temperature. 
Translating the density axis by $\phi_l$ facilitates comparing different systems.

\begin{figure*}[h]
 \centering
 \includegraphics[angle=0,width=0.85\textwidth]{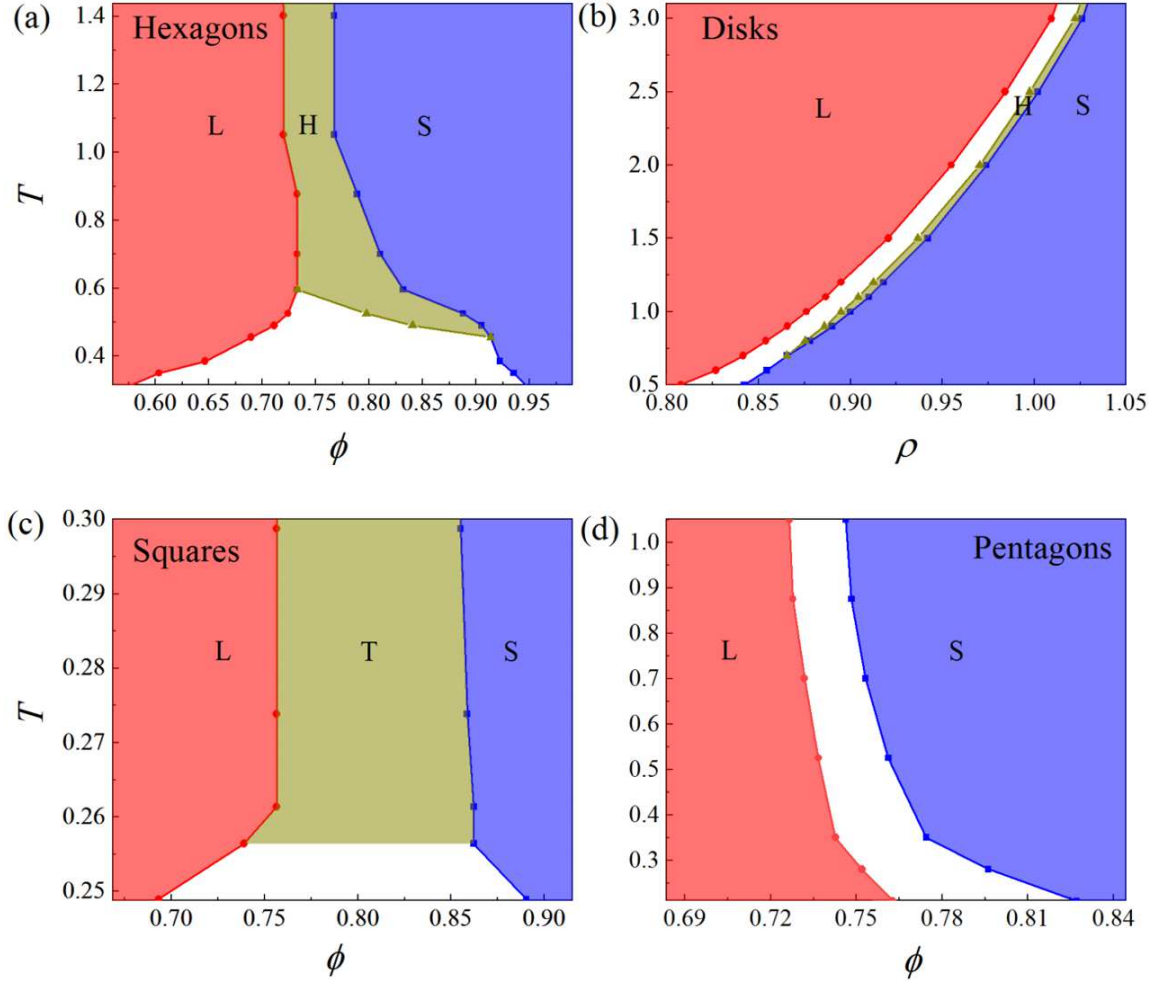}
\caption{
Phase diagrams of attractive polygons and LJ disks. 
The phase diagrams for (a) hexagons, (c) squares and (d) pentagons are plotted in the $T-\phi$ plane, and the one for (b) disks is in the $T-\rho$ plane.
Colours are used to distinguish the pure phases, liquid (L), solid (S), hexatic (H) and tetratic (T). Coexistence regions, including hexatic-liquid and solid-liquid coexistence, are white.
\label{fig:pd2}}
\end{figure*}

We present the same phase diagrams in the $T$ vs $\phi$ (or $\rho$) plane, to ease experimental comparisons, in Fig.~\ref{fig:pd2}.
The highest area fraction (or number density) of the liquid phase, and the lowest one of the solid one, have a system-specific temperature dependence.
In disks, both densities decrease with the temperature, while in pentagons the opposite occurs. In hexagons and squares, the maximum liquid area fraction decreases while the minimum solid area fraction increases.

\end{document}